\documentclass[useAMS,usenatbib]{mn2e}
\usepackage{graphicx}
\usepackage{txfonts}
\usepackage{natbib}
\usepackage{hyperref}
\usepackage{color}

\title[A polarized view of NGC~4151]
      {The polarized spectral energy distribution of NGC~4151}
\author[Marin et al.]
      {F.~Marin$^1$\thanks{E-mail: frederic.marin@astro.unistra.fr},
       J.~Le~Cam$^2$, E. Lopez-Rodriguez$^3$, M.~Kolehmainen$^1$, 
       B.~L.~Babler$^4$ and M.~R.~Meade$^4$\\
       $^1$Universit\'e de Strasbourg, CNRS, Observatoire Astronomique de Strasbourg, UMR 7550, F-67000 Strasbourg, France\\
       $^2$Institut d'optique Graduate School, Palaiseau, France\\
       $^3$SOFIA Science Center, NASA Ames Research Center, Moffett Field, CA 94035, USA\\
       $^4$Department of Astronomy, University of Wisconsin-Madison, Madison, WI, USA}

\date{Accepted 2020 May 25.
      Received 2020 May 15;
      in original form 2020 March 17.}

\pagerange{\pageref{firstpage}--\pageref{lastpage}}
\pubyear{2020}

\begin{document}

\maketitle

\begin{abstract}
NGC~4151 is among the most well-studied Seyfert galaxies that 
does not suffer from strong obscuration along the observer's line-of-sight. This 
allows to probe the central active galactic nucleus (AGN) engine with photometry, spectroscopy, reverberation mapping or interferometry. 
Yet, the broadband polarization from NGC~4151 has been poorly examined in the past 
despite the fact that polarimetry gives us a much cleaner view of the AGN physics than 
photometry or spectroscopy alone. In this paper, we compile the 0.15 -- 89.0~$\mu$m 
total and polarized fluxes of NGC~4151 from archival and new data in order to examine 
the physical processes at work in the heart of this AGN. We demonstrate that, from 
the optical to the near-infrared (IR) band, the polarized spectrum of NGC~4151 shows a much bluer 
power-law spectral index than that of the total flux, corroborating the presence of 
an optically thick, locally heated accretion flow, at least in its near-IR emitting radii.
Specific signatures from the atmosphere of the accretion structure are tentatively found 
at the shortest ultraviolet (UV) wavelengths, before the onset of absorption opacity. 
Otherwise, dust scattering appears to be the dominant contributor from the near-UV to 
near-IR polarized spectrum, superimposed onto a weaker electron component. We also 
identify a change in the polarization processes from the near-IR to the mid-IR, most
likely associated with the transition from Mie scattering to dichroic absorption 
from aligned dust grains in the dusty torus or narrow-line region. Finally, we 
present and discuss the very first far-infrared polarization measurement of NGC~4151 
at 89~$\mu$m.
\end{abstract}


\label{firstpage}

%________________________________________________________________

\section{Introduction}
\label{Introduction}
NGC~4151 holds a particular place in the field of active galactic nuclei (AGN). It was among the original list of the six ``\textit{extragalactic 
nebulae with high-excitation nuclear emission lines superposed on a normal G-type spectrum}'' observed by Carl K. Seyfert, that later gave his name 
to a specific class of AGNs \citep{Seyfert1943}. NGC~4151 is also one of the nearest galaxies to Earth, $z \sim$ 0.00332, which corresponds to a 
Hubble distance of 18.31 $\pm$ 1.31~Mpc \citep{Vaucouleurs1991}. Its optical and ultraviolet (UV) fluxes are known to vary quite dramatically, resulting 
in a fluctuating optical-type classification. NGC~4151 changes from a Seyfert-1.5 type in the maximum activity state to a Seyfert-1.8 in the minimum 
state \citep{Antonucci1983,Shapovalova2008}. Indeed, long-term temporal behaviors of this AGN showed flux variation from a factor 2 to 6 over a period of 12 years 
\citep{Ulrich1991,Shapovalova2008}. The X-ray flux of NGC~4151 is also known to be variable but is also very bright, explaining why it was one of the 
first Seyfert galaxies to be detected in the high energy sky by the Uhuru Satellite \citep{Gursky1971,Warwick1995}. A complete 0.1 -- 100~keV coverage 
of the source showed that the X-ray spectrum of NGC~4151 is a complex association of emitting and absorbing components arising from various locations, 
from the central engine to the extended polar outflows of the AGN \citep[e.g.,][]{Schurch2002,Schurch2004}. This complexity underlies a 
fundamental question: are all AGNs such convoluted systems or is NGC~4151 unique? In order to better understand the intrinsic physics and structure
of Seyfert galaxies, it becomes of great interest to determine whether this Seyfert galaxy is an archetype of its class or very far from it 
\citep{Zdziarski2002}.

The proximity of NGC~4151 allowed to measure several of its prime attributes. The narrow-line region (NLR) of this AGN has an inclination close 
to 45$^\circ$ \citep{Fischer2013,Fischer2014,Marin2016} with the north side out of the plane of sky away from our line-of-sight (LOS), 
which makes NGC~4151 an intermediate type Seyfert galaxy. This means that we are able to directly see the central engine (a supermassive black hole 
and its associated putative accretion disk) through the dust funnel of the circumnuclear gas and dust distribution that characterizes AGNs (see the 
fundamental paper on the subject by \citealt{Antonucci1993}). This dusty equatorial obscurer is the reason why we cannot see the central engine in type-2
AGNs where the observer's LOS intercepts this optically thick medium. The possibility to peer the central engine in NGC~4151 allowed 
to estimate its black hole mass to be of the order of 4 $\times$ 10$^7$~M$_\odot$ \citep{Bentz2006,Onken2007}. The supermassive black hole appears to be 
maximally spinning \citep{Cackett2014} and its inferred mass accretion rate is about 0.013~M$_\odot$ yr$^{-1}$ \citep{Crenshaw2007}. The putative 
geometrically thin, optically thick, accretion disk that surrounds the black hole is responsible for the multi-temperature thermal emission that 
produces an UV-to-optical signature called the Big Blue Bump. At the end of the accretion disk lies the broad emission line region (BELR) 
that is responsible for the detected broad emission lines in the UV, optical and near-infrared (near-IR) spectrum of AGNs \citep{Gaskell2009}. The 
size of the BELR is object-dependent but it is generally admitted that it lies between the accretion disk and the inner radius of the dusty circumnuclear
region. In NGC~4151, the radius at which dust grains start to survive the intense radiation field was measured thanks to reverberation mapping studies
and is of the order of 0.04~pc \citep{Minezaki2004}. Dust is responsible for another detectable feature in the spectral energy distribution (SED) 
of AGN, usually referred to as the IR bump and most likely associated with re-radiated emission from the dust grains. 

The superposition of the emission lines and the IR re-emission onto the UV/optical continuum makes it difficult to precisely measure 
the shape and the peak of the Big Blue Bump. This is unfortunate since the Big Blue Bump provides critical information on the structure and 
condition of the innermost AGN components. In order to isolate the true SED of the central component, it is advised to look at the polarized 
light of the AGN. Indeed, the polarized flux shaves off the unpolarized line emission from the extended AGN polar outflows or the host galaxy 
\citep{Kishimoto2004,Kishimoto2008}. In addition, polarimetry can detect variations in the emission/scattering physics thanks to two additional
and independent parameters: the polarization degree and polarization position angle \citep{Kishimoto2008,Marin2018}. In this paper, we thus 
investigate the polarized light of NGC~4151 thanks to archival polarimetric campaigns and compile, for the first time, its UV-to-IR polarized SED.
We aim at characterizing the emission from the Big Blue Bump, i.e. from the accretion disk itself, by eliminating the parasitic light from the 
BELR and from the dusty components in the near-IR. Additionally, we want to determine what are the dominant mechanisms that are responsible 
for the observed polarization from the UV to the far-IR band, allowing us to build a better picture of the central AGN components. The paper 
is organized as follows: Section \ref{Spectrum} describes the archive data and examines the total and polarized SED, which are further discussed 
in Section \ref{Discussion}. In Section \ref{Conclusions}, we present the conclusions.

%//////////////////////////////////////////////////////////////////
\section{The polarized flux spectrum of NGC~4151}
\label{Spectrum}

\subsection{Compiling the data}
\label{Spectrum:data}

\begin{table*}
  \centering
\begin{tabular}{ p{4cm} p{6.8cm} p{3.5cm} p{1.7cm} }
  \hline
  \textbf{Reference} & \textbf{Instrument} & \textbf{Waveband ($\mu$m)} & \textbf{Aperture ($"$)} \\
  \hline
    \textit{MAST archives} & Wisconsin Ultraviolet Photo Polarimeter Experiment (WUPPE) & 0.15 -- 0.32 & 4.2$"$ \\ 
    \citet{Thompson1979} & Steward Observatory 2.25 m telescope using the UCSD 200-channel Digicon & 0.32 -- 0.36 & 2.25$"$ \\
    \citet{Schmidt1980} & Lick Observatory (3m Shane telescope) & 0.37 -- 0.71 & 2.8$"$ \\
    \citet{Axon1994} & 4.5m William Herschel Telescope & 0.48 -- 0.75 & ? \\
    \citet{Kruszewski1977} & Steward 229 cm, Catalina 154 cm, Mt. Lemmon 102 cm reflectors & R (0.86), V (0.55), G (0.52), B (0.44), U (0.36),N (0.33) & 10$"$ \\
    \citet{Ruiz2003} & dual-beam polarimeter IRPOL2 at the United Kingdom Infrared Telescope (UKIRT) & J(1.23), H(1.64) & 1$"$ \\
    \citet{Kemp1977} & Steward 2.25 m telescope & 2.22 & 7.8$"$ \\
    \citet{Lopez2018} & CanariCam on the 10.4-m Gran Telescopio CANARIAS & 8 -- 13 & 0.4$"$ \\
    Lopez-Rodriguez et al. (in prep.) & HAWC+ onboard the 2.7-m Stratospheric Observatory for Infrared Astronomy (SOFIA) & 89 & 7.80"\\
  \hline
\end{tabular}
  \caption{Catalog of published, archival (WUPPE) and new (SOFIA) polarimetric measurements of 
	  NGC~4151. The first column is the reference paper, the second column is 
	  the instrument used for the measurement, the third column is the waveband
	  or filters used during the observation and the fourth column is the observation
	  aperture (in arcseconds). The exact instrument aperture is not know in the 
	  case of \citet{Axon1994}'s publication but it is likely to be of the order 
	  of 4$"$, see the optical spectropolarimetry of 3C~234 taken with the same 
	  instrument by \citet{Young1998}.}
  \label{Tab:Catalog}
\end{table*}

\begin{figure}
  \centering
  \includegraphics[trim = 0mm 0mm 0mm 0mm, clip, width=8.8cm]{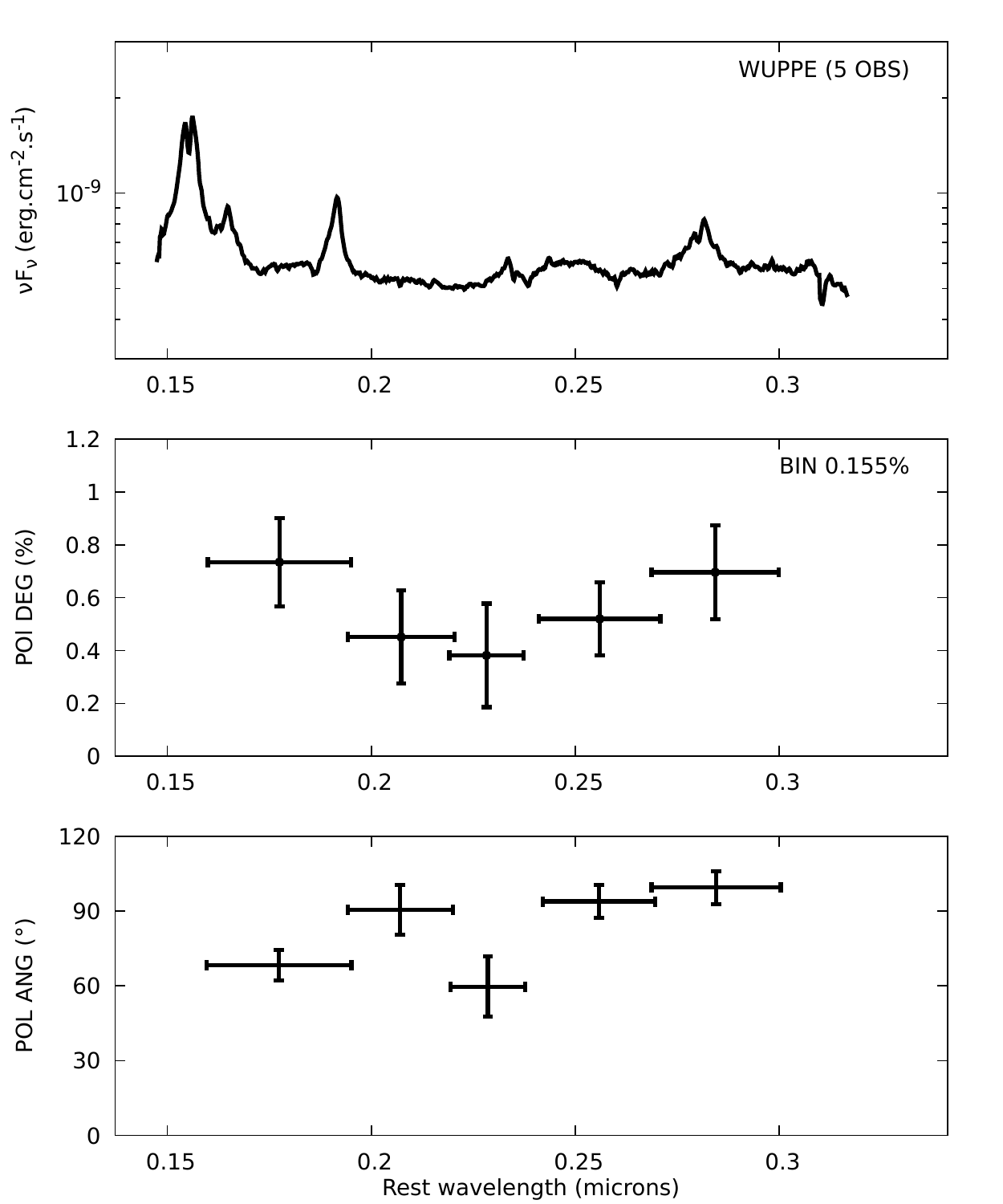}
  \caption{Unpublished MAST archival compilation of all the polarimetric 
	  observations of NGC~4151 made with WUPPE. It sums up 5 observation
	  campaigns taken between March 4$^{th}$ and March 13$^{th}$ 1995 
	  for a total of 3136 seconds.}
  \label{Fig:WUPPE}
\end{figure}

We skimmed through the SAO/NASA Astrophysics Data System (ADS) digital library and gathered all the publications reporting polarization 
measurements in NGC~4151. We found 19 papers spanning from 1971 to 2018. Two papers reported circular polarization measurements 
\citep{Nikulin1971,Angel1976} and, since we are interested in linear polarization only (the thermally-emitted Big Blue Bump should not 
produce intrinsically circularly polarized light), we safely discarded them\footnote{The optical circular polarization measurement 
of \citet{Nikulin1971} has been contradicted by \citet{Gehrels1972}}. Due to the diversity of instruments, the polarimetric data 
are also spanning from apertures that widely vary from sub-arcsecond scales to more than 40 arcseconds. It is obvious that we cannot reconstruct
a polarized SED using dramatically different apertures so we focused on apertures lower than 10$"$. At an heliocentric distance of 18.31~Mpc,
this corresponds to a linear size of 0.88~kpc. Since AGN polar outflows are often optically detected up to a projected distance of 1~kpc 
(see, e.g., \citealt{Schmitt1996}), we ensure that the polarized SED mostly accounts for the AGN flux, not the host galaxy. Finally, we
favored publications with spectropolarimetric data instead of measurements taken with broad filters. This allows us to have a much more 
precise reconstruction of the shape of the Big Blue Bump. We ended up with seven publications covering as many wavelengths as possible 
from the optical to the near/mid IR (see Tab.~\ref{Tab:Catalog}). To account for the UV band, we retrieved the old and unpublished 
polarization measurements taken by the Wisconsin Ultraviolet Photo Polarimeter Experiment (WUPPE, \citealt{Code1993,Nordsieck1994}). 
Five polarimetric observations in the 0.15 -- 0.32~$\mu$m band were obtained with WUPPE and compiled by the WUPPE team into a 3136 seconds
long exposure spectrum visible in Fig.~\ref{Fig:WUPPE}. The observations were temporally close enough that variability was not an issue 
for stacking the Stokes parameters. We have also included newly acquired imaging polarimetric data at 89 $\mu$m~using the High-resolution
Airborne Wideband Camera/Polarimeter (HAWC+) onboard the 2.7-m NASA's Stratospheric Observatory for Infrared Astronomy (SOFIA). These 
observations are part of an AGN polarimetric survey at far-IR wavelengths under the program 07\textbackslash0032 (PI: Lopez-Rodriguez).
Observations were performed on January the 28$^{th}$, 2020, with a total on-source time of 3360 seconds and will be presented in detail
in a follow-up manuscript (Lopez-Rodriguez et al. in prep.). For the goal of this project we performed aperture photometry within the 
beam size of 7\farcs80 (0.69 kpc) at 89 $\mu$m~and estimated that the nucleus of NGC 4151 is unpolarized (0.9$\pm$0.8\%) with an 
undetermined position angle of polarization. The final compilation of published, archival and new polarimetric measurements of NGC~4151
is presented in Tab.~\ref{Tab:Catalog}. The total and polarized fluxes will be presented in Fig.~\ref{Fig:Spectrum} while the polarization
degrees and position angles will be shown in Fig.~\ref{Fig:POL_angle}. 

%//////////////////////////////////////////////////////////////////
\subsection{Examination of the total and polarized fluxes}
\label{Spectrum:examination}

\begin{figure*}
  \centering
  \includegraphics[trim = 0mm 0mm 0mm 0mm, clip, width=18cm]{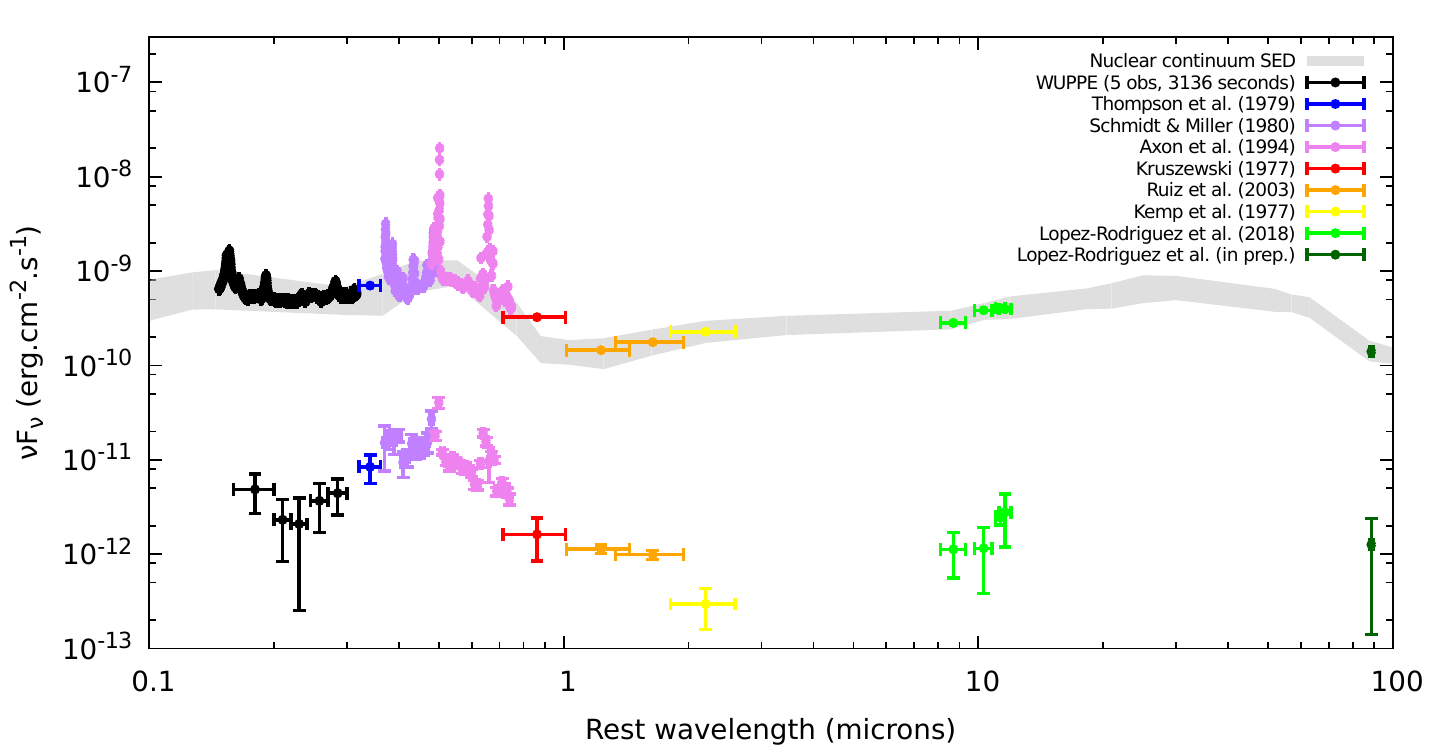}
  \caption{Broadband 0.15 -- 89.0~$\mu$m total flux (top) and 
	    polarized flux (bottom) spectra of NGC~4151 measured 
	    from various instruments (see Tab.~\ref{Tab:Catalog}).
	    The gray region represents the averaged nuclear 
	    continuum SED extracted from the NASA/IPAC Extragalactic 
	    Database.}
  \label{Fig:Spectrum}
\end{figure*}

\begin{figure}
  \centering
  \includegraphics[trim = 0mm 0mm 0mm 0mm, clip, width=8.7cm]{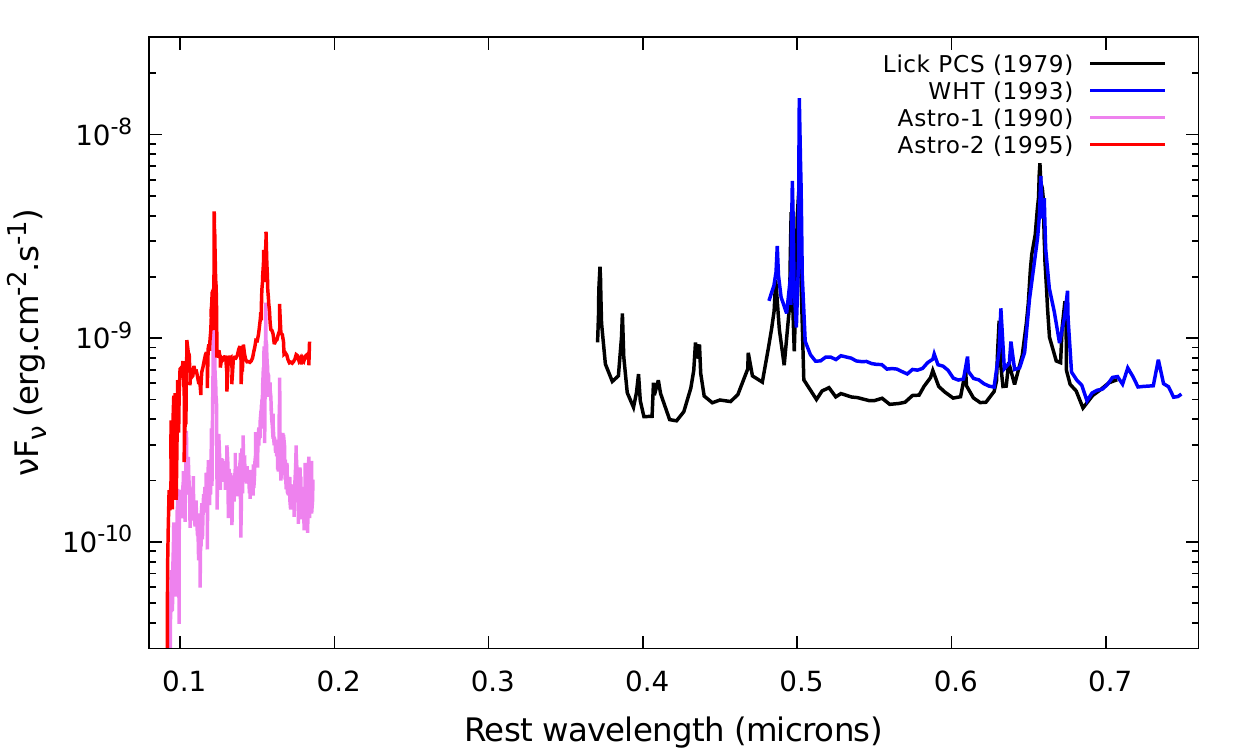}
  \caption{Example of the strong UV and optical variability 
	   in NGC~4151. The far-UV fluxes are from the Hopkins 
	   Ultraviolet Telescope on Astro-1 \citep{Kriss1992} 
	   and Astro-2 \citep{Kriss1995}, while the optical
	   fluxes are from the Pockels cell spectropolarimeter (PCS) 
	   of Lick Observatory \citep{Schmidt1980} and the 
	   4.5m William Herschel Telescope (WHT) by \citet{Axon1994}.}
  \label{Fig:Variability}
\end{figure}

We present in Fig.~\ref{Fig:Spectrum} the broadband, 0.15 -- 89.0~$\mu$m, total flux (top) and polarized flux (bottom) spectra of 
NGC~4151. The total flux is entirely coherent with past measurements \citep{Kriss1995,Alexander1999} and with the nuclear SED retrieved 
from the NASA/IPAC Extragalactic Database. The nuclear SED, shown in shaded gray in Fig.~\ref{Fig:Spectrum}, is the averaged SED
compiled from continuum flux measurements with apertures lower than 10$"$. The thickness of the gray area corresponds to the observed 
flux fluctuations (including the error bars) and the missing data were linearly interpolated between two contiguous points.
We note that due to the variable nature of NGC~4151 in the UV and optical bands, and since it is a composite spectrum aggregating
about 40 years of measurements, it is not gainful to fit the total flux data points in order to obtain a representative spectral index for the 
underlying continuum. Although the intensity spectrum is coherent with the nuclear data averaged from NED, it has been shown that the 
continuum can vary by a factor of 6 in less than 10 years, impeding a trustworthy estimation of the spectral index from the optical to
near-IR waveband ($\approx$ 0.5 -- 1~$\mu$m). We illustrate this in Fig.~\ref{Fig:Variability}, where we can see that the spectral index 
of the UV and optical/near-IR power-laws underlying the continuum are drastically different due to variability. Fortunately, in the 
near-to-mid IR, where the accretion disk spectrum starts to be hidden under the hot dust thermal emission from the equatorial dusty region, 
variability is less of an issue (see Fig.~\ref{Fig:Variability}). This allowed us to fit the IR (1 -- 12~$\mu$m) spectral index
from the total flux SED: F$_\nu \propto \nu^{-1.33}$. This spectral index is entirely consistent with the average value of $-1.48$ $\pm$ 0.30 
reported by \citet{Alonso2003} for the IR (1 -- 16~$\mu$m) spectral index of a sample of 22 Seyfert 1--1.5 galaxies.

To unveil the signature of the external parts of the putative accretion disk that should produce a much bluer spectral index than the
total flux from the optical to the near-IR, we plot the polarized flux of NGC~4151 in Fig.~\ref{Fig:Spectrum} (bottom spectrum). The data 
are entirely consistent with partially-to-unpolarized line emissions superposed on a smooth polarized continuum \citep{Schmidt1980}. 
The optical-to-near-IR spectral slope is clearly more pronounced (bluer) in polarized flux and extends up to 2.2~$\mu$m.
Yet again, the non-simultaneity of the optical and near-IR observations makes it unproductive to measure a spectral index. However, 
it has been demonstrated by \citet{Gaskell2012} that the polarized flux of NGC~4151 follows the total flux with a lag of a few days,
allowing us to correlate the two quantities in order to suppress the variable component. This permits us to provide a qualitative
measurement of the difference between the spectral index of the total and polarized fluxes. From power least squares fittings, we
find a difference of $\sim$0.6. This is entirely consistent with \citet{Kishimoto2008}, in which a optical-to-near-IR power-law spectral
index difference of $\sim$0.74 between the total and polarized fluxes of their quasar sample can be estimated from their Fig.~1. Although
we cannot precisely measure the optical-to-near-IR spectral slope in NGC~4151 using a multi-epoch composite spectrum, we confirm the 
methodology of \citet{Kishimoto2008}: the polarized light spectrum of NGC~4151 seems to corroborate the existence of an optically thick, 
thermally heated accretion disk structure, at least in its outer near-IR emitting radii. Single epoch, multi-wavelength polarimetric
data are required to get a quantitative conclusion. 

\subsubsection{Polarization degree and angle}
\label{Spectrum:polarization}

\begin{figure*}
  \centering
  \includegraphics[trim = 0mm 0mm 0mm 0mm, clip, width=17.8cm]{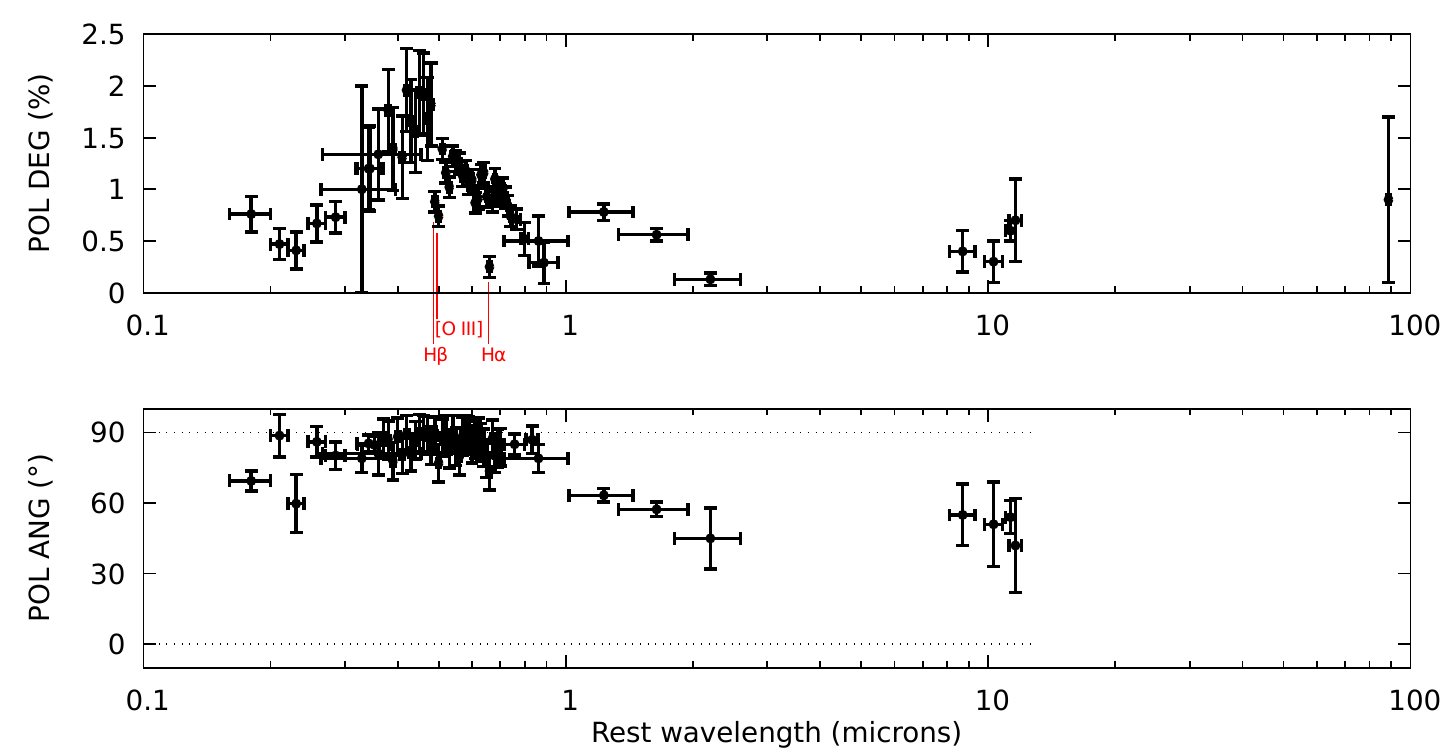}
  \caption{Broadband polarization degree (top) and polarization 
	    angle (bottom) of NGC~4151. The polarization angle
	    was reduced to the interval 0 -- 90$^\circ$ (see 
	    references in Tab.~\ref{Tab:Catalog}). The polarization
	    angle at 89~$\mu$m is undefined because the core is 
	    consistent with an unpolarized source.}
  \label{Fig:POL_angle}
\end{figure*}

Looking at the polarization degree of NGC~4151 in Fig.~\ref{Fig:POL_angle}, we observe a rather wavelength-dependent spectrum. 
We first made sure that the polarization degree and angle continua were not subject to variability due to the different epochs
of observation. \citet{Gaskell2012} demonstrated that the B-band degree and angle of polarization of NGC~4151 did not vary (were 
constant within the observational error bars) between 1997 and 2003. A similar conclusion can be drawn between 0.48 and 0.71~$\mu$m,
were the polarized observations of \citet{Schmidt1980} and \citet{Axon1994} indicate that the degree and angle of polarization 
remained constant over $\sim$ 15 years. We are thus sure that intrinsic variability, such as synchrotron polarization induced in a 
failed jet, is not an issue here. The polarization we report has been corrected for instrumental polarization by the respective 
authors and interstellar polarization is not an issue. NGC~4151 is situated close to the north galactic pole (RA: 182$^\circ$.635745, Dec: 39$^\circ$.405730) so the 
interstellar contribution is negligible. It was measured to be 0.07\% close to 0.44~$\mu$m by \citet{Kruszewski1977}. The only 
remaining source of extra-nuclear light is the host galaxy, a (R')SAB(rs)ab galaxy type according to \citet{Vaucouleurs1991}. 
The starlight of such spiral galaxy usually peaks in the 0.7 -- 2~$\mu$m band (in $\nu$F$_\nu$, \citealt{Polletta2007}) and, in 
the case of NGC~4151, the AGN light should dominate shortward of 0.5~$\mu$m \citep{Schmidt1980}. So, where does the strong
wavelength-dependence of the UV polarization come from? If electron scattering was the sole physical mechanism responsible 
for the observed polarization, both the polarization degree and angle would have been constant from the UV to the near-IR. 
Here, the increase of the polarization degree shortward 0.23~$\mu$m is qualitatively explained by the accretion disk atmosphere 
models presented by \citet{Blaes1996}. In order to explain and reproduce far-UV polarimetric observations of AGNs, the authors
developed a numerical tool to calculate fully self-consistent models of pure-hydrogen disk atmospheres. They included the effects
of opacity absorption (both from free-free absorption and photo-ionization) and found that UV polarization from disk atmospheres 
rises at shorter wavelengths. A maximum is reached between 0.1 and 0.2~$\mu$m, where a rotation on the polarization position 
angle might occur, depending on the effective temperature and surface gravity of the model (see Figs.~2 and 3 in \citealt{Blaes1996}). 
Below the Lyman edge (0.0912~$\mu$m) the polarization is expected to drop dramatically before rising up to several percents
in the extreme-UV. Unfortunately, there is no polarimetric data on NGC~4151 below 0.16~$\mu$m so a far-UV spectropolarimeter 
is needed to confirm that the rise of polarization we see at 0.16 -- 0.2~$\mu$m in Fig.~\ref{Fig:POL_angle} is real. 

From 0.2 to 0.5~$\mu$m the polarization degree rises, from $\sim$0.5\% to $\sim$2.0\%. This wavelength-dependence is the signature 
of dust scattering, most likely from the inner surfaces of the dusty circumnuclear region and, less likely, from the dust grains 
inside the polar regions. The polarization position angle is close to 90$^\circ$ which, when compared to the 450~milliarseconds 
radio structure along a position angle $\approx$ 83$^{\circ}$ observed with VLBI/Merlin \citep{Harrison1986}, indicates that 
equatorial scattering dominates. Indeed, subtracting the radio position angle from the measured polarization angle gives a result 
close to 0$^\circ$. This means that the polarization position angle is parallel to the small-scale radio axis, hence scattering 
mainly occurs along the equatorial plane and not inside the polar outflows. It was shown by \citet{Goosmann2007} that an isolated 
dusty torus, viewed at an inclination close to its half-opening angle is able to produce a polarization degree that rises at longer 
wavelengths, up to several percents. A more complex structure inclined at 45$^\circ$, including an accretion disk, a BELR, a 
circumnuclear obscurer and ionized polar winds is also capable of producing a polarization degree that increases with increasing 
wavelengths, together with a parallel polarization position angle \citep{Marin2012}. Mie scattering is thus the dominant mechanism 
producing the 1 -- 10$"$ near-UV/optical polarization in NGC~4151. Electron scattering also occurs but is less dominant, otherwise 
the polarization degree would be grayer, such as seen in the first arc-second around the core of NGC~1068 \citep{Miller1991,Antonucci1994,Marin2018}. 
Interestingly, our result differs from the conclusions of \citet{Kishimoto2008} who found that electron scattering is
the dominant scattering mechanism in their sample. This difference is actually due to target selection: NGC~4151 is a low-luminosity
AGN (Seyfert galaxy) while \citet{Kishimoto2008}'s sample consists of 6 high-luminosity AGNs (quasars). Because quasars are much 
brighter than Seyferts, radiative pressure tends to destroy a larger fraction of dust in the close environment of the AGN, resulting in 
circumnuclear dusty regions that are much thinner (geometrically flattened along the equatorial plane) than in the Seyfert's case
\citep{Lawrence1982,Sazonov2015,Marin2016}. In particular, \citet{Sazonov2015} found that low-luminosity AGNs tend to have more
massive torus, situated closer to the central engine, than quasars. This results in a larger solid angle for light-dust interaction 
at the inner radius of the circumnuclear dust in NGC~4151 than for quasars, hence a stronger impact of Mie scattering onto the 
observed polarization. Our results conveniently confirm that low-luminosity AGNs are likely to be embedded in a dustier environment
than quasars.

From 0.5~$\mu$m to 1.0~$\mu$m, the polarization degree decreases due to the onset of the flux dominance from the host galaxy 
\citep{Schmidt1980}. The dilution is due to unpolarized starlight that peaks in this regime. The same behavior was observed
in the polarized spectrum of NGC~1068 \citep{Marin2018}. The polarization angle remains constant, indicating that equatorial 
scattering still dominates from 0.5 to 1.0~$\mu$m. The polarization position angle only deviates from the parallel 
alignment after 1.0~$\mu$m. The polarization angle rotates from $\sim$83$^\circ$ to 45$^\circ$ at 2.2~$\mu$m then stabilizes up 
to the mid-IR (8 -- 12~$\mu$m). This variation is of particular interest because it clearly tells us that the physical mechanisms
responsible for the production of polarization in the IR band in NGC~4151 is no longer dust scattering, as in the optical. 
Such finding must be put in the light of an infrared study of the polarization in NGC~4151 by \citet{Lopez2018}. In their work, 
the authors stated ``\textit{The nearly constant degree and [angle] of polarization from 1~$\mu$m to 12~$\mu$m strongly suggest 
that a single polarization mechanism dominates in the core of NGC~4151}''. We entirely agree with them. However, they also 
stated that ``\textit{scattering off optically thick dust ... is consistent with the observed nearly constant degree and 
[polarization angle]}''. This is a reasonable assumption only if the 1 -- 12~$\mu$m band is isolated from the optical and UV 
polarization data. In our extended study, we showed that dust scattering occurs in the optical but is no longer the main 
mechanism producing polarization in the infrared due to the rotation of the polarization angle\footnote{\citet{Lopez2018} did 
not account for the infrared polarization measurement obtained by \citet{Kemp1977} at 2.2~$\mu$m that would have otherwise 
better highlighted the smooth variation of the polarization angle in the near-infrared band in their study.}. Another supporting
clue for this statement is the fact that the bump in polarization around 1 -- 2~$\mu$m cannot be produced by Mie scattering alone 
\citep{Young1995,Efstathiou1997}. In fact, looking at Fig.~3 from \citet{Lopez2018}, dichroic absorption from aligned dust 
grains appears to be a better explanation for the wavelength-dependent IR polarization in NGC~4151. This is also supported by 
the simulations from \citet{Young1995}, where a dichroic component is necessary to produce the expected polarization degrees 
longward of 1.5~$\mu$m (see their Fig.6, bottom panel). In-between 12 and 89~$\mu$m, another change in the polarization process
happens, since polarization from magnetically aligned dust grains dominates over other polarization mechanisms (e.g. dust/electron
scattering, dichroic absorption, and synchrotron emission). The null polarization at 89~$\mu$m can then arise from 1) a dominant
turbulent magnetic field in the central tens parsecs from the core, and/or 2) a depolarization effect due to the large beam 
(7.80$"$, 0.68~kpc), averaging a more complex underlying field.

%//////////////////////////////////////////////////////////////////
\section{Discussion}
\label{Discussion}

The polarimetric data from WUPPE must be taken with caution. In the UV regime, polarimetric measurements of AGN ionizing fluxes 
have a low signal-to-noise ratio. For this reason, unbinned polarization plots always appear to show greater polarization than 
binned results. This is due to the fact that polarization is a positive quantity. When calculating the average polarization over 
a wavelength range, the mean values of the Stokes parameters Q and U are error-weighted but WUPPE suffered from instrumental 
issues that could not be reliably calibrated. For this reason there is no data between 2368 and 2430~\AA, and the measurements 
in the 1600 -- 3000~\AA~band should be analyzed with circumspection. The error bars in Fig.~\ref{Fig:WUPPE} are 1-$\sigma$ errors
and the Q and U values are in different quadrants of the Q-U plane, which partially explains why the polarization angle may 
disagree so much between two consecutive bins (e.g., the 1600 -- 2000 and the 2000 - 2200~\AA~bins). Finally, there is a 
correction to the polarimetric value that is not the simple error weighted means of $\sqrt{\langle Q \rangle^2 + \langle U \rangle^2)}$. 
The adjustment of the final polarimetric values as a function of polarimetric signal-to-noise is described in \citet{Nordsieck1976}. 

It was shown by \citet{Impey1995} and \citet{Koratkar1995} that at least 3 high-redshift AGNs have a
remarkably high polarization in the far-UV. One of the best examples is PG~1630+377 ($z \approx$ 1.48). Its UV polarization 
degree rises rapidly up to 20\% below the Lyman edge, a value that is usually found in blazing sources. Numerous scenarios 
for explaining such polarization features has been proposed by \citet{Koratkar1995}, including Faraday screens covering 
highly polarized continuum sources and geometric dilution. We now know that source obscuration is not an option as the 
inclination of NGC~4151 is such that we can directly see its central engine, allowing reverberation mapping studies
\citep[e.g.][]{Bentz2006,Gaskell2012}. Simulations by \citet{Blaes1996} also failed to reproduce such dramatic value using
a pure-hydrogen disk atmosphere model. More recently, new computations by \citet{Chang2017} tried to compute the Rayleigh
polarization emerging from scattered radiation around the Lyman alpha wavelength in thick neutral regions but they could
not apply their results to PG~1630+377. One of the reasons for such slow progresses is the lack of data. We only have a 
handful of UV polarimetric measurements from high redshift quasars ($z >$ 0.5). 

The 1 -- 12~$\mu$m polarization of 6 Seyfert galaxies shows a diverse set of physical processes arising from its cores 
\citep{Lopez2018}. For radio-quiet obscured AGN, the polarization is mainly arising from dichroic absorption by the dusty
obscured surrounding the active nuclei or host galaxy. However, for NGC 4151, an un-obscured radio-quiet AGN, the physical
component producing the IR polarization is difficult to identify. The reason is due to the competing polarization mechanisms 
of the several physical components within the beam size of the observations. We here provide a potential interpretation based 
on the shape of the 1--89 $\mu$m polarized spectrum and radiative alignment torque mechanism by \citet{Lee2019}. These authors
show that the polarization peak arising from dichroic absorption by aligned dust grains changes as a function of the radiation 
field, extinction, and dust grain properties. To reproduce our 1 -- 2~$\mu$m bump polarization, a direct view to the source 
with low extinction, A$_{v} \le 5$, oblate grains with axial ratio of 1.5, and dust grains expose to weak radiation field 
are required \citep[Fig. 7 in][]{Lee2019}. This physical environment predicts a negligible emissive polarization at 89~$\mu$m
with a rise of polarization at longer wavelengths \citep[Fig. 10 in][]{Lee2019}. We interpret these results as that our 
1 -- 89~$\mu$m polarization arises from elongated and aligned dust absorption in the 1 -- 12~$\mu$m wavelength range and emission 
at 89~$\mu$m. As the dust grains are exposed to weak radiation field, the dust may be located in the outer areas of the torus 
or NLR as suggested by \citet{Lopez2018}. Our approach has shown that a detailed analysis of the total and polarized flux, 
the degree and position angle of polarization, together with high-angular resolution observations, is a powerful tool to 
identify the most likely mechanisms producing polarization in the whole electromagnetic spectrum.

%//////////////////////////////////////////////////////////////////
\section{Conclusions}
\label{Conclusions}

We have compiled, for the first time, the polarized SED of NGC~4151, one of the archetypal pole-on Seyfert galaxies. 
Despite the inherent problem of UV and optical variability, we have demonstrated that the optical-to-near-IR power-law
spectral index of NGC~4151 is much bluer in polarized flux. This is in perfect agreement with past studies of 
quasars polarized SED, strengthening the case for the use of polarimetry to unveil the true nature of the Big Blue Bump.
We have analyzed the wavelength-dependent polarization of this object and found a potential signature of the accretion 
disk atmosphere in the shortest wavelengths. However, due to the unreliability of the instrument in this waveband, 
new UV polarimetric observations are necessary. We have also established that the continuum polarization of NGC~4151
is mainly due to dust scattering from the near-UV to the near-IR. Compared to the sample of \citet{Kishimoto2008}, 
who found that electron scattering dominates in higher-luminosity AGNs, our result tend to confirm that low-luminosity
objects such as NGC~4151 are embedded in dustier environments. A smooth rotation of the polarization position angle 
between 1 and 2.2~$\mu$m identified a change in the polarization mechanism that we attribute to the onset of dichroic 
absorption from aligned dust grains. In the far-IR, polarization from dichroic absorption becomes less dominant and 
polarized emission from elongated and aligned dust grains prevails. Our results demonstrate that single epoch, broadband, 
polarimetric measurements are necessary to unveil the physical mechanisms that produce polarization in the heart of AGNs.
We thus advocate for new polarimeters that could cover the a large fraction of the UV-to-IR electromagnetic spectrum.

In addition, WUPPE data are the only spectropolarimetric UV measurements of NGC~4151 so far. This pull the trigger 
on the importance of new UV spectropolarimeters in the future. If we aim at investigating the UV properties of nearby 
AGNs, it is necessary to probe the shortest wavelengths in order to uncover the potential signatures from the accretion 
disk or its atmosphere. Spectropolarimetric observations of nearby AGN at rest wavelengths below the Lyman edge are thus 
mandatory to push forward the analysis. Future spatial instruments such as POLLUX, a high-resolution UV spectropolarimeter 
proposed for the 15-meter primary mirror option of LUVOIR \citep{Bouret2018,Marin2019}, or satellites at higher energies 
such as IXPE \citep{Weisskopf2016,Marin2017} are necessary to shed light on this topic. Furthermore, high-spatial 
resolution observations in the 1-- 200 $\mu$m can provide us with detailed information about the physical scales of the 
dusty structure obscuring the AGN as well as the presence/absent of magnetic fields in the accretion flow to the active 
nuclei. Future ground-based 30-m class telescopes and space-base telescopes equipped with sensitive polarimeters, such 
as SPICA \citep{Adami2019} and Origins \citep{Staguhn2018}, are thus also mandatory.

\section*{Acknowledgments}
We would like to thank the anonymous referee for taking the time to write an encouraging report on our paper.
The authors are grateful to Robert ``Ski'' Antonucci for his remarks and suggestions that improved the quality of 
this article. FM would like to thank the Centre national d'\'etudes spatiales (CNES) who funded his post-doctoral grant 
``Probing the geometry and physics of active galactic nuclei with ultraviolet and X-ray polarized radiative
transfer''. Based partially on observations made with the NASA/DLR Stratospheric Observatory for Infrared Astronomy (SOFIA)
under the 07\textbackslash0032 Program. SOFIA is jointly operated by the Universities Space Research Association, Inc. (USRA), under
NASA contract NAS2-97001, and the Deutsches SOFIA Institut (DSI) under DLR contract 50 OK 0901 to the University of Stuttgart.

\bibliographystyle{mnras}
\bibliography{biblio}

\label{lastpage}

\end{document}